\RequirePackage{ifpdf}
\ifpdf 
\documentclass[pdftex]{sigma}
\else
\documentclass{sigma}
\fi

\begin{document}
\allowdisplaybreaks

\renewcommand{\PaperNumber}{054}

\FirstPageHeading

\ShortArticleName{Geodesic Flow and Two (Super) Component Analog
of the Camassa--Holm  Equation}

\ArticleName{Geodesic Flow and Two (Super) Component Analog \\
of the Camassa--Holm  Equation}

\Author{Partha GUHA~$^\dag$ and Peter J. OLVER~$^\ddag$}

\AuthorNameForHeading{P. Guha and P.J. Olver}

\Address{$^\dag$~S.N. Bose National Centre for Basic Sciences,
JD Block, Sector-3, Salt Lake,\\
$\phantom{\dag}$~Calcutta-700098, India}
\EmailD{\href{mailto:partha@bose.res.in}{partha@bose.res.in}}

\Address{$^\ddag$~School of Mathematics, University of Minnesota,
Minneapolis, MN 55455, USA}
\EmailD{\href{mailto:olver@math.umn.edu}{olver@math.umn.edu}}
\URLaddressD{\url{http://www.math.umn.edu/~olver/}}

\ArticleDates{Received March 08, 2006, in f\/inal form May 08,
2006; Published online May 23, 2006}

\Abstract{We derive the $2$-component Camassa--Holm equation and
corresponding $N=1$ super generalization as geodesic f\/lows with
respect to the $H^1$ metric on the extended Bott--Virasoro and
superconformal groups, respectively.}


\Classification{53A07; 53B50}

\newcommand{\p}{\partial}
\newcommand{\g}\gh
\newcommand{\bo}{\ltimes}
\newcommand{\rd}{\partial}
\newcommand{\J}{J}

\def\fr#1#2{{\textstyle {#1\over #2}}}
\def\ip#1#2{\langle \,{#1}\,,\,{#2}\, \rangle}
\def\Ip#1#2{\left \langle \,{#1}\,,\,{#2}\,\right \rangle}

\def\bbk#1{\bigl [\,{#1}\,\bigr ]}
\def\Pa#1{\left (\,{#1}\,\right )}

\def\g{\mathfrak g} \def\gh{\widehat \g}
\def\ga{\gh^{\,\ast}}
\def\greg{\ga_{\rm reg}}

\section{Introduction}

About ten years ago, Rosenau, \cite{ro}, introduced a class of
solitary waves with compact support as solutions of certain wave
equations with nonlinear dispersion. It was found that the
solutions of such systems unchanged from collision and were thus
called {\it compactons}. The discovery that solitons may
compactify under nonlinear dispersion inspired further
investigation of the role of nonlinear dispersion. It has been
known for some time that nonlinear dispersion causes wave breaking
or lead to the formation of corners or cusps. Beyond compactons, a
wide variety of other exotic non-analytic solutions, including
peakons, cuspon, mesaons, etc., have been found in to exist in a
variety of models that incorporate nonlinear
dispersion~\cite{LOR}.

We will study integrable evolution equations appearing in
bi-Hamiltonian form
\begin{gather}\label{1}
u_t = \J_1 \frac{\delta H_{1}}{\delta u} = \J_2 \frac{\delta
H_0}{\delta u}, \qquad n = 0,1,2, \ldots ,
\end{gather}
where $J_1$ and $J_2$ are compatible Hamiltonian operators.  The
initial Hamiltonians $H_0$, $H_1$ are the f\/irst two in a
hierarchy of
 conservation laws
 whose corresponding bi-Hamiltonian f\/lows are successively generated by the
recursion operator
\[
{\cal R} =  \J_2\J_{1}^{-1}.
\]
We refer the reader to \cite{O} for the basic facts about
bi-Hamiltonian systems.

In an earlier work, the second author showed with Rosenau
\cite{or} that a simple scaling argument shows that most
integrable bi-Hamiltonian systems are governed by tri-Hamiltonian
structures. They formulated a method of ``tri-Hamiltonian
duality'', in which a recombination of the Hamiltonian operators
leads to integrable hierarchies endowed with nonlinear dispersion
that supports compactons or peakons. A related construction can be
found in the contemporaneous paper of Fuchssteiner \cite{FuchCH}.

The tri-Hamiltonian  formalism can be best described through
examples. The Korteweg--de~Vries equation
\begin{gather*}
u_t = u_{xxx} + 3uu_x,
\end{gather*}
can be written in bi-Hamiltonian form \eqref{1} using the two
compatible Hamiltonian operators
\[
\J_1 = D, \qquad \J_2 = D^3 + uD + Du, \qquad \mbox{where} \quad
D \equiv \frac{d}{dx}
\]
and
\[
H_1 = \frac{1}{2}\int u^2 \, dx, \qquad H_2 =  \frac{1}{2}\int
\left(-u_{x}^{2} + u^3 \right)\, dx.
\]

The tri-Hamiltonian duality construction is implemented as
follows:
\begin{itemize}
\itemsep=0pt \item A simple scaling argument shows that $\J_2$ is
in fact the sum of two compatible Hamiltonian operators, namely
$K_2 = D^3$ and $K_3 = uD + Du$, so that $K_1 = J_1,K_2,K_3$ form
a triple of mutually compatible Hamiltonian operators.

\item Thus, when we can recombine the Hamiltonian triple  by
transferring the leading term~$D^3$ from $\J_2$ to $\J_1$, thereby
constructing the Hamiltonian pairs $\widehat{\J_1} = K_2 \pm K_1 =
D^3 \pm D$. The resulting self-adjoint operator $S = 1 \pm D^2$ is
used to def\/ine the new f\/ield variable $\rho = Su = u \pm
u_{xx}$.

\item Finally, the second Hamiltonian structure is constructed by
replacing $u$ by $\rho$ in the remaining part of the original
Hamiltonian operator $K_3$, so that $\widehat{\J_2} = \rho D +
D\rho$.  Note that this change of variables does not af\/fect
$\widehat{\J_1}$.
\end{itemize}

As a result of this procedure, we recover the tri-Hamiltonian dual
of the KdV equation
\begin{gather}\label{3}
\rho_t = \widehat{\J_1}\frac{\delta {\widehat H}_2}{\delta \rho} =
\widehat{\J_2}\frac{\delta {\widehat H}_1}{\delta \rho},
\end{gather}
where
\[
{\widehat H}_1 = \frac{1}{2}\int u\rho \, dx = \frac{1}{2}\int
\left(u^2 \mp u_{x}^{2}\right) \, dx, \qquad {\widehat H}_2 =
\frac{1}{2}\int \left(u^3 \mp uu_{x}^{2}\right) \, dx.
\]
In this case, \eqref{3} reduces to the celebrated Camassa--Holm
equation \cite{ch1,ch2}:
 \begin{gather*}
u_t \pm u_{xxt} = 3uu_x \pm \left(uu_{xx} +
\fr{1}{2}u_{x}^{2}\right)_x.
\end{gather*}

\begin{remark}
The choice of plus sign leads to an integrable equation which
supports compactons, whereas the minus sign is the water wave
model derived by Camassa--Holm, whose solitary wave solutions have
a sharp corner at the crest.
\end{remark}

{\bf The Ito equation.} Let us next study the Ito equation
\cite{ito},
\begin{gather*}
u_t = u_{xxx} + 3uu_x + vv_x , \nonumber \\
v_t = (uv)_x ,
\end{gather*}
which is a prototypical example of a two-component KdV equation.
This can also be expressed in bi-Hamiltonian form using the
following two Hamiltonian operators
\[
\J_1 = \begin{pmatrix}
                    D & 0 \\
                            0 & D \\
                      \end{pmatrix},
\qquad \J_2 = \begin{pmatrix}
                     D^3 + uD + Du & vD \\
                            Dv & 0 \\
                      \end{pmatrix},
\]
with Hamiltonians
\[
H_1 = \frac{1}{2}\int \left(u^2 + v^2\right) \, dx, \qquad H_2 =
\frac{1}{2}\int\left(u^3 + uv^2 - u_{x}^{2}\right) \, dx.
\]
Again, a simple scaling argument is used to split
\[
\J_2 = \begin{pmatrix}
                    D^3 & 0 \\
                            0 & 0 \\
                      \end{pmatrix}
                      + \begin{pmatrix}
                    uD + Du & vD \\
                            Dv & 0 \\
                      \end{pmatrix}
\]
 as a sum of two compatible Hamiltonian operators.
To construct the dual, we transfer the leading term $D^3$ from the
f\/irst Hamiltonian operator to the second. We obtain the f\/irst
Hamiltonian operator for the new equation
\[
\widehat{\J_1} = \begin{pmatrix}
                    D \pm D^3 & 0 \\
                            0 & D \\
                      \end{pmatrix} \equiv D \begin{pmatrix}
                     S & 0 \\
                            0 & 1 \\
                      \end{pmatrix}.
\]
Therefore, the new variables are def\/ined as
\[
\begin{pmatrix}
                     \rho \\
                            \sigma \\
                      \end{pmatrix} =  \begin{pmatrix}
                    S & 0 \\
                            0 & 1 \\
                      \end{pmatrix}\begin{pmatrix}
                     u \\
                            v \\
                      \end{pmatrix}.
\]
The second Hamiltonian structure follows from the truncated  part
of the original Hamiltonian operator $\J_2$, so that
\[
\widehat{\J_2} = \begin{pmatrix}
                    \rho D + D\rho & vD \\
                            Dv & 0 \\
                      \end{pmatrix}
\]
with
\[
{\widehat H}_1 = \frac{1}{2}\int \big(u\rho + v^2\big) \, dx,
\qquad {\widehat H}_2 = \frac{1}{2}\int \big(u^3 + uv^2 \mp
uu_{x}^{2}\big) \, dx.
\]
The dual system \eqref{3} takes the explicit form
\begin{gather*}
u_t \pm u_{xxt} = 3uu_x + vv_x +
\left( uu_{xx} + \fr{1}{2}u_{x}^{2}\right )_x, \nonumber \\
v_t = (uv)_x .
\end{gather*}

\noindent {\bf Motivation.} The Camassa--Holm equation was derived
physically as a shallow water wave equation by Camassa and Holm
\cite{ch1,ch2,hmr}, and identif\/ied as the geodesic f\/low on the
group of one-dimensional volume-preserving dif\/feomorphisms under
the $H^1$ metric.  The multi-dimensional analogs lead to important
alternative models to the classical Euler equations of f\/luid
mechanics. Later, Misiolek \cite{mi} showed that, like the KdV
equation \cite{ok}, it can also be characterized as a~geodesic
f\/low on the Bott--Virasoro group.

\medskip

Recently, a $2$-component generalization of the Camassa--Holm
equation has drawn a lot of interest among researchers. Chen,
Dubrovin, Falqui, Grava, Liu and Zhang (the group at SISSA) have
been working on multi-component analogues, using reciprocal
transformations and studying their ef\/fect on the Hamiltonian
structures, \cite{clz,fal,lz}. They show that the 2-component
system cited above admits peakons, albeit of a dif\/ferent shape
owing to the dif\/ference in the corresponding
 Green's functions.
Another two-component generalization also appeared recently as the
bosonic sector of the extended $N=2$ supersymmetric Camassa--Holm
equation \cite{popo}.

 Following Ebin--Marsden \cite{ebma}, we enlarge ${\rm Diff}\,(S^1)$ to a Hilbert manifold
${\rm Diff}^s (S^1)$, the dif\/feo\-mor\-phisms of the Sobolev
class $H^s$. This is a topological space. If $s > n/2$, it makes
sense to talk about an $H^s$ map from one manifold to another.
Using local charts, one can check whether the derivations of order
$ \leq s$ are square integrable. The Lie algebra of ${\rm
Diff}^s(S^1)$ is denoted by ${\rm Vect}^s(S^1)$.

In this paper we show that a $2$-component generalization of the
Camassa--Holm equation
 and its super analog also follow from the geodesic with respect to the
 $H^1$ metric on the semidirect product space
 ${\rm Diff}^s(S^1) \ltimes C^{\infty}(S^1)$ and its supergroup respectively.
In fact, it is known that numerous coupled KdV equations
\cite{gu1,gu2,gu3} follow from geodesic f\/lows of the right
invariant~$L^2$ metric on the semidirect product group ${\widehat
{{\rm Diff}\,(S^1) \ltimes C^{\infty}(S^1)}}$~\cite{ackp, mor}.

\section{Preliminaries}

The Lie algebra of ${\rm Diff}^s(S^1) \bo C^{\infty}(S^1)$ is the
semidirect product Lie algebra
\[
 \g = {\rm Vect}^s(S^1) \bo C^{\infty}(S^1).
\]
An element of $\g$ is a pair
\[
\left(f(x)\frac{d}{dx}, a(x)\right),\qquad {\rm where} \qquad
f(x)\frac{d}{dx} \in {\rm Vect}^s(S^1), \qquad {\rm and} \qquad
a(x) \in C^{\infty}(S^1).
\]
It is known that this algebra has a three dimensional central
extension given by the non-trivial cocycles
\begin{gather}
\omega_1 \left(\left(f(x)\frac{d}{dx}, a(x)\right),
\left(g\frac{d}{dx}, b\right)\right) =
\int_{S^1}f^{\prime}(x)g^{\prime \prime}(x)\,dx,\nonumber\\
\omega_2 \left(\left(f(x)\frac{d}{dx}, a(x)\right),
\left(g\frac{d}{dx}, b\right)\right) =
\int_{S^1} \bbk{f^{\prime \prime}(x)b(x) - g^{\prime \prime}(x)a(x)}\, dx,\label{7}\\
\omega_3 \left(\left(f(x)\frac{d}{dx}, a(x)\right),
\left(g\frac{d}{dx}, b\right)\right)=
2\int_{S^1}a(x)b^{\prime}(x)\, dx.\nonumber
\end{gather}
 The f\/irst cocycle $\omega_1$ is the well-known
Gelfand--Fuchs cocycle. The Virasoro algebra
\[
{\rm Vir} = {\rm Vect}^s(S^1) \oplus {\mathbb R}
\]
is the unique non-trivial central extension of ${\rm Vect}^s(S^1)$
based on the Gelfand--Fuchs cocycle. The space $C^{\infty}(S^1)
\oplus {\mathbb R}$ is identif\/ied as {\it regular part} of the
dual space to the Virasoro algebra. The pairing between this space
and the Virasoro algebra is given by:
\[
\Ip{ (u(x), a)}{\left(f(x)\frac{d}{dx}, a(x)\right)} =
\int_{S^1}u(x)f(x)\,dx + a\alpha.
\]

Similarly we consider the following extension of $\g$,
\[
\gh  = {\rm Vect}^s(S^1) \bo C^{\infty}(S^1) \oplus {\mathbb R}^3.
\]
The commutation relation in $\gh $ is given by
\[
\left[\left(f\frac{d}{dx}, a, \alpha\right) , \left(g\frac{d}{dx},
b, \beta\right)\right]  := \left((fg^{\prime} -
f^{\prime}g)\frac{d}{dx}, fb^{\prime} - ga^{\prime},
\omega\right),
\]
where $\alpha = (\alpha_1, \alpha_2, \alpha_3)$, $\beta =
(\beta_1, \beta_2, \beta_3) \in {\mathbb R}^3$, and where $\omega
= (\omega_1, \omega_2, \omega_3)$ are the cocycles.

Let
\[
\greg = C^{\infty}(S^1) \oplus C^{\infty}(S^1) \oplus {\mathbb
R}^3
\]
denote the {\it regular part} of the dual space $\ga$ to the Lie
algebra $\gh$, under the following pairing:
\[
\ip {\widehat u}{\widehat f} = \int_{S^1}\bbk{f(x)u(x)+a(x)v(x)}\,
dx + \alpha \cdot \gamma,
\]
where ${\widehat u} = ( u(x), v, \gamma) \in \greg$, ${\widehat f}
= \left( f\frac{d}{dx}, a, \alpha\right)\in \gh$. Of particular
interest are the coadjoint orbits in $\greg$. In this case,
Gelfand, Vershik and Graev~\cite{GVG}, have constructed some of
the corresponding representations.

Let us introduce $H^1$ inner product on the algebra $\gh$
\[
\ip{\widehat f}{\widehat g}_{H^1}
 = \int_{S^1}\bbk{f(x)g(x)+a(x)b(x)+\rd_x{f(x)}\rd_x{g(x)}}\,dx +
\alpha \cdot \beta,
\]
where
\[
{\widehat f} = \left( f\frac{d}{dx}, a, \alpha\right), \qquad
{\widehat g} = \left( g\frac{d}{dx}, b, \beta\right).
\]
Now we compute:

\begin{lemma}
The coadjoint operator with respect to the $H^1$ inner product is
given by
\[
{\rm ad}^{\ast}_{\hat f}\,\left(\begin{array}{c}
u \\
v
 \end{array} \right) =
\left( \begin{array}{c} (1- {\partial}^2)^{-1}[2f^{\prime}(x)(1 -
\rd_{x}^{2})u(x) + f(x) (1 - \rd_{x}^{2})u^{\prime}(x) +
a^{\prime}v(x)]\vspace{1mm}\\
f^{\prime}v(x) + f(x)v^{\prime}(x)
\end{array}  \right).
\]
\end{lemma}

\begin{proof} Since we have identif\/ied ${\frak g}$ with ${\frak g}^{\ast}$,
 it follows from the def\/inition that
\begin{gather*}
\ip { {\rm ad}^{\ast}_{\hat f}\, {\widehat u}}{\widehat g}_{H^1}
 = \ip { {\widehat u}}{[{\widehat f},\widehat g]}_{H^1}\\
 \phantom{\ip { ad^{\ast}_{\hat f}\, {\widehat u}}{\widehat g}_{H^1}}{}
 = - \int_{S^1} \bbk{(fg^{\prime} - f^{\prime}g)u -
(fb^{\prime} - ga^{\prime})v
 - \rd_x(fg^{\prime} - f^{\prime}g)\rd_xu} dx.
\end{gather*}
After computing all the terms by integrating by parts and using
the fact that the functions $f(x)$, $g(x)$, $u(x)$ and $a(x)$,
$b(x)$, $v(x)$ are periodic, the right hand side can be expressed
as above.

Let us compute now the left hand side:
\begin{gather*}
{\rm ad}^{\ast}_{\hat f}\,\left(\begin{array}{c}
u \\
v
 \end{array} \right) = \int_{S^1} \bbk{({\rm ad}^{\ast}_{\hat f}\,u) g  +
({\rm ad}^{\ast}_{\hat f}\,u)^{\prime}g^{\prime} + ({\rm
ad}^{\ast}_{\hat f}\,v)b }\,dx
\\
\phantom{{\rm ad}^{\ast}_{\hat f}\,\left(\begin{array}{c}
u \\
v
 \end{array} \right)}{} = \int_{S^1} \bbk{[(1- {\partial}^2){\rm ad}^{\ast}_{\hat f}\, u]g +
 ({\rm ad}^{\ast}_{\hat f}\,v)b }\, dx
= \big\langle ((1- {\partial}^2){\rm ad}^{\ast}_{\hat f}\,u ,
({\rm ad}^{\ast}_{\hat f}\,v)), (g,b)\big\rangle.
\end{gather*}
Thus by equating the the right and left hand sides, we obtain the
desired formula.
\end{proof}

We conclude that the Hamiltonian operator arising from the induced
Lie--Poisson structure is
\[
\left(\begin{array}{cc}
D\rho  + \rho D  & vD\\
Dv  & 0
\end{array} \right),
\]
where $\rho  = ( 1 - \p_{x}^2)u$.  We conclude that

\begin{theorem*}
A curve
\[
{\widehat c}(t) = \Pa{u(x,t)\frac{d}{dx}, v(x,t), \gamma } \subset
\g
\]
defines a geodesic in the
 $H^1$ metric if and only if
\begin{gather*}
u_{t} - u_{xxt} = 3uu_x + vv_{x}
- \Pa{uu_{xx} + \fr12u_{x}^{2}}_x, \\
v_{t} = 2(uv)_x.
\end{gather*}
\end{theorem*}

\section[Geodesic flow and superintegrable systems]{Geodesic f\/low and superintegrable systems}

The f\/irst and foremost characteristic property of a superalgebra
is that all the additive groups of its basic and derived
structures are ${\mathbb Z} _2$ graded. A {\em vector superspace}
is a ${\mathbb Z} _2$ graded vector space $V = V_B \oplus V_F$.
An element $v$ of $V_B$ (resp. $V_F$) is said to be even or
bosonic (resp. odd or fermionic). The super-commutator of a pair
of elements $v, w \in V$ is def\/ined to be the element
\[
[ v , w ] = vw - (-1)^{{\bar v}{\bar w}} wv.
\]

The generalized Neveu--Schwartz superalgebra \cite{ov} is composed
of two parts: the bosonic (even) and the fermionic (odd). These
are given by
\[
S\g_B = {\rm Vect}^s(S^1) \oplus C^{\infty}(S^1) \oplus {\mathbb
R} ^3, \qquad S\g_F = C^{\infty}(S^1) \oplus C^{\infty}(S^1).
\]
There are {\it three} dif\/ferent actions:
\begin{enumerate}\itemsep=0pt

\item[(A)] the action of the bosonic part on the bosonic part,
discussed earlier. \item[(B)] the action of the bosonic part on
the fermionic part, given by
\begin{gather*}
[~,~] : S\g_B \otimes S\g_F \longrightarrow S\g_F,
\\
[ (f(x)\frac{d}{dx}, a(x)), (\phi(x), \alpha(x)) ] :=
 \left(\begin{array}{c}
f(x)\phi^{\prime} - \frac{1}{2}f^{\prime}(x)\phi(x) \vspace{2mm}\\
f(x)\alpha^{\prime}(x) + \frac{1}{2}f^{\prime}(x)\alpha(x) -
\frac{1}{2}a^{\prime}(x)\phi(x)
\end{array}\right) .
\end{gather*}
\item[(C)] the action of   the fermionic part on   the fermionic
part, given by
\begin{gather*}
[~,~]_+ : S\g_F \otimes S\g_F \longrightarrow S\g_B,
\\
[(\phi(x),\alpha(x)),(\psi(x),\beta(x))]_+ =
\left(\phi\psi\frac{d}{dx}, \phi \beta + \alpha \psi,
\omega_F\right),
\end{gather*}
where $\omega_F = (\omega_{F1},\omega_{F2},\omega_{F3})$ is the
fermionic cocycle, with components
\begin{gather*}
\omega_{F1} ((\phi,\alpha), (\psi,\beta)) =
2\int_{S^1}\phi^{\prime}(x)\psi^{\prime}(x)\,dx,
\\
\omega_{F2} ((\phi,\alpha), (\psi,\beta)) = -2\int_{S^1}
(\phi^{\prime}(x)\beta(x) + \psi^{\prime}\alpha(x))\,dx,
\\
\omega_{F3} ((\phi,\alpha), (\psi,\beta)) =
4\int_{S^1}\alpha(x)\beta(x)\,dx.
\end{gather*}
The supercocycle $\omega_S$ has two parts, the bosonic and the
fermionic:
\[
 \omega_S = \omega_B \oplus \omega_F,
 \]
where the bosonic part $\omega_B$ is identical to $\omega =
(\omega _1, \omega _ 2, \omega _ 3)$, as given  by \eqref{7}.
\end{enumerate}

With this in hand, we establish the supersymmetric $2$-component
generalization of the Ca\-mas\-sa--Holm equation.

\begin{definition*}
The $H^1$ pairing between the regular part of the dual space
$S\ga$ and $S\g$ is given by
\begin{gather*}
\Ip{(u(x), v(x), \psi(x), \beta)}
{(f(x)\frac{d}{dx}, a(x), \phi(x), \alpha)}_{H^1} \\
\qquad{} =\int_{S^1} f(x)u(x)\, dx + \int_{S^1}f_xu_x \, dx +
\int_{S^1} a(x)v(x) \, dx  \\
\qquad \quad {} + \int_{S^1}\phi(x)\psi(x)\, dx +
\int_{S^1}\phi_x\psi_x \, dx + \int_{S^1} \alpha(x)\beta(x)\, dx .
\end{gather*}
\end{definition*}

Let us compute the coadjoint action with respect to the $H^1$
norm.

\begin{lemma}
\begin{gather*}
{\rm ad}^{\ast}_{\hat f}\,\left(\begin{array}{c}
u(x) \\
v(x) \\
\psi(x)\\
\beta(x)
\end{array}\right)  \\
 = \left(\begin{array}{@{}c@{}}
(1 - \partial^2)^{-1}[2f^{\prime}(1 - \partial^2)u(x) + (1 -
\partial^2)u^{\prime}f + a^{\prime}v
+ \frac{1}{2}(1 - \partial^2)\psi^{\prime}\phi + \frac{3}{2}(1 - \partial^2)\psi\phi^{\prime}]\vspace{1mm}\\
f^{\prime}v + fv^{\prime} + \frac{1}{2}(\beta^{\prime}\phi + \beta \phi^{\prime})\vspace{1mm}\\
(1 - \partial^2)^{-1}[f(1 - \partial^2)\psi^{\prime} +
\frac{3}{2}f^{\prime}(1 - \partial^2)\psi
+ \frac{1}{2}a^{\prime}\beta + (1 - \partial^2)u\phi + v\alpha] \vspace{1mm}\\
f\beta^{\prime} + \frac{1}{2}f^{\prime}\beta + v\phi
\end{array}\right) .
\end{gather*}
\end{lemma}

\begin{proof}[Sketch of the proof.]
Using the def\/inition of the coadjoint action
\[
\ip{{\rm ad}^{\ast}_{\hat f}\,{\widehat u}}{\widehat g}_{H^1} =
\ip{\widehat f}{[{\widehat u}, {\widehat g}] }_{H^1}
\]
with
\[
{\widehat f} =
 \left(\begin{array}{c}
f(x) \\
a(x) \\
\phi(x)\\
\alpha(x)
\end{array}\right),
\qquad {\widehat u} =  \left(\begin{array}{c}
u(x) \\
v(x) \\
\psi(x)\\
\beta(x)
\end{array}\right),
\qquad {\widehat g} =  \left(\begin{array}{c}
g(x) \\
b(x) \\
\chi(x)\\
\gamma(x)
\end{array}\right),
\]
we obtain
\[
\Ip{(u, v, \psi, \beta)]}{\left(\begin{array}{c}
(fg^{\prime} - f^{\prime}g)\frac{d}{dx} + \phi\chi \frac{d}{dx} \vspace{1mm}\\
fb^{\prime} - ga^{\prime} + \phi\gamma + \alpha\chi \vspace{1mm}\\
f\chi^{\prime} - \frac{1}{2}f^{\prime}\chi + g\phi^{\prime} -
\frac{1}{2}g^{\prime}\phi\vspace{1mm}\\
f\gamma^{\prime} + \frac{1}{2}f^{\prime}\gamma -
\frac{1}{2}a^{\prime}\gamma + g\alpha^{\prime} +
\frac{1}{2}g^{\prime}\alpha - \frac{1}{2}b^{\prime}\phi
\end{array}\right)}_{H^1}.
\]
This would give us the right hand side without the $(1 -
\partial^2)^{-1}$ term, which appears on the left hand side:
\begin{gather*}
{\rm L.H.S.} = \int_{S^1} ({\rm ad}^{\ast}_{\hat f}\,u) g \,dx +
\int_{S^1} ({\rm ad}^{\ast}_{\hat f}\,u)^{\prime}g^{\prime}\, dx
\int_{S^1} ({\rm ad}^{\ast}_{\hat f}\,v)b \,dx
\\
\phantom{{\rm L.H.S.} =}{} + \int_{S^1} ({\rm ad}^{\ast}_{\hat
f}\,\psi)\phi \, dx + \int_{S^1} ({\rm ad}^{\ast}_{\hat
f}\,\psi^{\prime})\phi^{\prime} \,dx + \int_{S^1} ({\rm
ad}^{\ast}_{\hat f}\,\beta)\alpha \, dx
\\
\phantom{{\rm L.H.S.}}{}=\int_{S^1} [(1- {\partial}^2){\rm
ad}^{\ast}_{\hat f}\,{u}]{g} \,
dx + \int_{S^1} ({\rm ad}^{\ast}_{\hat f}\,v)b \,dx\\
\phantom{{\rm L.H.S.} =}{} + \int_{S^1} [(1- {\partial}^2){\rm
ad}^{\ast}_{\hat f}\,{\psi}]{\phi} \,dx  + \int_{S^1} ({\rm
ad}^{\ast}_{\hat f}\,\beta)\alpha \,dx.
\end{gather*}
Equating the right and left hand sides, we obtain the desired
formula.
\end{proof}

Therefore, if we use the Euler--Poincar\'e equation and the
computational trick used in \cite{ds}, we obtain the
supersymmetric version of the two component Camassa--Holm
equation:
\begin{gather*}
m_t = 2mu_x + m_xu + (vv_x) + 3\xi \xi^{\prime \prime} ,
\\
v_t = 2(uv)_x + \beta^{\prime}\xi^{\prime} + \beta \xi^{\prime
\prime},
\\
(1 - \partial^2)\xi_t = 4m\xi^{\prime} + 3m^{\prime}\xi +
2\xi^{\prime \prime \prime},
\\
\beta_t = 2u\beta^{\prime} + u^{\prime}\beta + 2v\xi^{\prime},
\end{gather*}
where $m = u - u_{xx}.$

\begin{corollary}
\begin{gather*}
{\rm ad}^{\ast}_{\hat f}\,{\widehat u} = \left(\begin{array}{c}
2uf^{\prime}(x) + u^{\prime}f + a^{\prime}v   + f^{\prime \prime
\prime}
+ \frac{1}{2}\psi^{\prime}\phi + \frac{3}{2}\psi \phi^{\prime}\vspace{1mm}\\
f^{\prime}v + fv^{\prime} + \frac{1}{2}(\beta^{\prime}\phi +
\beta \phi^{\prime})\vspace{1mm}\\
f\psi^{\prime} + \frac{3}{2}f^{\prime}\psi +
\frac{1}{2}a^{\prime}\beta + u\phi + v\alpha
+ 2\phi^{\prime \prime}\vspace{1mm}\\
f\beta^{\prime} + \frac{1}{2}f^{\prime}\beta + v\phi
\end{array}\right) .
\end{gather*}
\end{corollary}

In this way, we recover a supersymmetric version of Ito equation
\cite{ito} given by
\begin{gather}
u_t = 6uu_x + 2(vv_x) + u_{xxx} + 3\xi \xi^{\prime \prime},
\nonumber\\
v_t = 2(uv)_x + \beta^{\prime}\xi^{\prime} + \beta \xi^{\prime
\prime},
\nonumber\\
\xi_t = 4u\xi^{\prime} + 3u^{\prime}\xi + 2\xi^{\prime \prime
\prime},
\nonumber\\
\beta_t = 2u\beta^{\prime} + u^{\prime}\beta +
2v\xi^{\prime}.\label{sIto}
\end{gather}

\begin{corollary}
In the supersymmetric Ito equation \eqref{sIto}:

{\rm (A)} if we set the super variables $\xi = \beta = 0$, we
recover the Ito equation.

{\rm (B)} If we set $v=\beta=0$,  we obtain
\begin{gather*}
u_t = 6uu_x +  u_{xxx} + 3\xi \xi^{\prime \prime},
\\
\xi_t = 4u\xi^{\prime} + 3u^{\prime}\xi + 2\xi^{\prime \prime
\prime},
\end{gather*}
which is a fermionic extension of KdV equation and, modulo
rescalings, is the super KdV equation of Mathieu and Manin--Radul
{\rm \cite{MMR1,MMR2}}.
\end{corollary}

\begin{remark} The physicists usually distinguish the fermionic and
supersymmetric extension among each other. From the physical point
of view the supersymmetry requires the invariance under the
supersymmetric transformations while for the fermionic extension
we have no such restriction. All extended equations  considered in
this paper are fermionic.
\end{remark}

\subsection*{Acknowledgements}
One of the authors (PG) would like to thank Professor Andy Hone for
various information about the two component Camassa--Holm
equation. PG is also grateful to Professor Valentin Ovsienko for
various stimulating discussions. PG would like to thank Professor
Dieter Mayer at TU Clausthal, where the part of work was done in a
stimulating atmosphere. This work was partially supported by the
DFG Research Group ``Zeta functions and locally symmetric spaces"
which is gratefully acknowledged.

The work of the second author was supported in part by NSF Grant
DMS 05--05293.

\LastPageEnding

\end{document}